# The unusual active galaxy H1821+643 and the elusive nature of FRI quasars

David Garofalo[1] & Chandra B. Singh[2]

1. Department of Physics, Kennesaw State University, Marietta GA 30060, USA, email: dgarofal@kennesaw.edu

2. South-Western Institute for Astronomy Research, Yunnan University, University Town, Chenggong, Kunming 650500, People's Republic of China, email: chandrasingh@ynu.edu.cn

Abstract

The moderate spin estimate for the black hole at the center of the cool core cluster H1821+643 motivates the completion of a story about this object's origin and evolution that was in the making since the work by Blundell & Rawlings over two decades ago as the first example of a massive black hole accreting at near Eddington rates with an FRI jet. This elusive combination of properties was explained in our 2010 model where we showed it to be part of a small parameter space that includes X-shaped radio galaxies. As an accreting black hole that never experienced a counterrotating phase, H1821+643 is constrained by theory to produce a jet for spin values $a$ satisfying $0.1 < a < \sim 0.7$ and an FRI jet for a slightly smaller range. The feedback from such a black hole is not subject to a tilted jet and is why star formation rates remain high in this cluster environment. The prediction is that H1821+643 is within millions of years of becoming jetless.

1. Introduction

Blundell & Rawlings (2001) discovered FRI quasars – near Eddington accreting supermassive black holes producing FRI jets and asked whether insufficient sensitivity was responsible for the failure to detect their alleged diffuse extended radio emission at higher redshifts. But two decades later and with deeper surveys, the detection of very few additional such objects has shown them to be rare. Why is that?

Optically, H1821+643 was discovered over three decades ago (Schneider et al 1992), and observed in X-rays shortly thereafter (Hall et al 1997; Saxton et al 1997). Although it was classified as a radio quiet quasar, Blundell & Rawlings (2001) discovered a jet long hundreds of thousands of pc. The nature of this type of object was explained/predicted by Garofalo, Evans & Sambruna (2010). In light of the recent black hole spin measurement of H1821+643 (Sisk-Reynes et al 2022), various properties can now be strung together in order to understand its origin and evolution. We show that while the parameter space for this kind of object is small, a subclass of active galactic nuclei (AGN) includes them, namely X-shaped radio galaxies. But FRI quasars can also form by themselves in mergers that lead to accretion around black holes whose spin is below about 0.7. Based on the observational signatures associated with relativistic reflection from the innermost stable orbit of disk accretion, Sisk-Reynes et al. (2022) concluded that the black hole in H1821+643 has spin values in the range of 0.25 – 0.84 on a 90% confidence level. In this work we try to establish the possible connection between evolution of black hole spin in the framework of the gap model (Garofalo, Evans & Sambruna 2010) with observational constraints of H1821+643. In Section 2 we describe the model prescription and the likely origin and evolution of H1821+643. In Section 3 we conclude.

2. The origin of FRI jets

Figures 1-3 capture the features of the model that allow us to understand where FRI quasars are directly visible, where they are hidden, and where they do not form at all. The first thing to note is that the FRI/FRII jet morphology dichotomy is prescribed to be associated with prograde/retrograde accretion onto rotating black holes (Garofalo, Evans & Sambruna, 2010). However, near-Eddington accretion may suppress jets for specific values of black hole spin. These are the key features needed to understand FRI quasars. Figure 1 shows them directly visible in the sense that will be understood in Figure 2 where they appear hidden. Figure 1 captures the evolution of black holes that are formed both in mergers and as a result of secular processes. In both cases, we imagine the black hole - or newly formed black hole – to have low or even zero spin. The cold gas accreting onto the black hole must spin it up and when the spin reaches about 0.2, the jet is powerful enough to be classified as an FRI (Garofalo & Singh 2019). As the black hole spins up, and the stability of circular orbits is affected by the dragging of spacetime, the inner edge of the disk moves closer to the black hole. We see this starting from the lower panel and moving upwards. This allows the system to tap more deeply into the gravitational potential of the black hole, which leads to more powerful radiatively driven disk winds as can be seen by the red arrows. Eventually, the disk wind is effective enough that it does not allow the inner disk material to load the magnetic field lines threading the inner disk. As a result, the Blandford-Payne (BP) jet cannot form. Since the overall jet is a combination of the BP jet and the Blandford-Znajek (BZ) jet from the black hole, the absence of the BP jet means that no jet is formed. This has been estimated to occur when the spin of the black hole is around 0.7 (Garofalo, Evans & Sambruna 2010; Garofalo & Singh 2016). As the jet turns off, the system becomes a jetless black hole accreting at near Eddington rates. This is referred to as a radio quiet quasar.

Figure 2 applies to merging isolated galaxies whose new black hole ends up spinning rapidly but such that the cold gas funneled into the black hole region settles into a counterrotating accretion disk. Such configurations are a minority because counterrotation is unlikely (King et al 2005; Garofalo, Christian & Jones 2019). Since corotation dominates, most post-merger field environments will lead to corotation around spinning black holes. Those scenarios are captured in Figure 1. The subset that end up in counterrotation, on the other hand, generate an FRII jet as shown in the lower panel of Figure 2. This is due to the strong BZ and BP effects based on the large value of the disk inner edge (Garofalo, Evans & Sambruna 2010). As in Figure 1, the cold gas is associated with high excitation and we have an FRII HERG. Counterrotation spins the black hole down to zero spin and then up again. Once a spin value of 0.2 is reached we are in the same state as the bottom panel of Figure 1, thus an FRI quasar is formed, but with a twist. The mass added to the black hole by accretion is given by (Raine & Thomas 2005),

$$\Delta m = \int dm(1-2m/3r_{ms})^{-0.5}$$

where m is the mass of the black hole and $r_{ms}$ is the marginally stable inner orbit of the accretion disk in units where Newton's constant and the speed of light are equal to 1. Mass adds angular momentum to the black hole which spins the hole down during the counterrotation phase but spins it up during the corotation phase. Given that about a 10% increase in mass is needed to change the black hole spin from -0.2 (retrograde) to 0.2 (prograde), we can estimate the time it takes to accomplish that at the Eddington accretion limit. This is about $5 \times 10^6$ years. As a result, the relic FRII-like jet (at the spin value of -0.2) is still visible. However, the transition through zero spin has likely changed the direction of the jet (Garofalo, Joshi et al 2020). In short, the FRI quasar is observed in tandem with a relic jet at some angle. Such objects are known as X-shaped radio galaxies. A few hundred such objects have been recorded. We refer to them here as hidden FRI quasars. The same evolution applies as in Figure 1 in that once the threshold spin of about 0.7 is achieved, jet suppression sets in. The top panel of Figure 2, in fact, shows a high spinning black hole with no jet.

Figure 3 is the only figure without FRI quasars, hence the labeling. Because the mergers that formed counterrotating accretion configurations were around the most massive black holes, the jet feedback was most dominant, and this led to a feedback effect that altered the state of accretion. Cold, radiatively efficient thin disks, evolve into advection dominated accretion flows (ADAF). This happens quickly as seen by the second to the lowest panel in Figure 3. The ADAF accretion state is associated with weak emission lines hence low excitation so the system is labeled an FRII LERG. It is clear that from this point on that no FRI quasar can emerge because the state of accretion does not change (i.e. there is no way to transition from ADAF to a thin disk). The characteristic end state of Figure 3 objects is a massive elliptical low excitation FRI radio galaxy such as M87. Note, finally, the persistent high star formation rate in Figure 1 compared to the suppression of the star formation rate in the prograde regime shown in Figures 2 and 3. This is the Roy Conjecture (Garofalo & Mountrichas 2022 and references therein) according to which tilted FRI jets are effective in heating the ISM and shutting down star formation. Since Figure 1 black holes did not experience a counterrotating phase, they did not transition through zero spin and therefore did not experience a new plane for accretion (Garofalo, Joshi et al 2020).

# Origin and evolution of FRI quasars

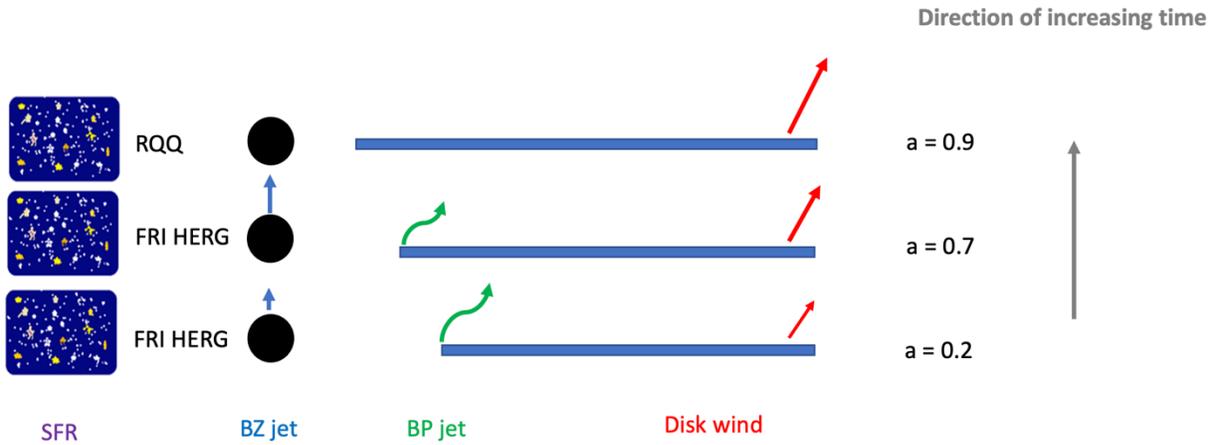

Figure 1: The origin and evolution of FRI quasars formed in mergers or in systems where secular processes feed the black hole. Left column: Star formation rate. Second column: jet morphology and excitation state with HERG = high excitation radio galaxy. Blue arrows: Strength of BZ jet. Green arrows: Strength of BP jet. Red arrows: Strength of radiative disk wind. Column labeled $a$: Black hole spin value. As the system accretes beyond a spin of 0.7, jet suppression sets in.

# Origin and evolution of X-shaped radio galaxies

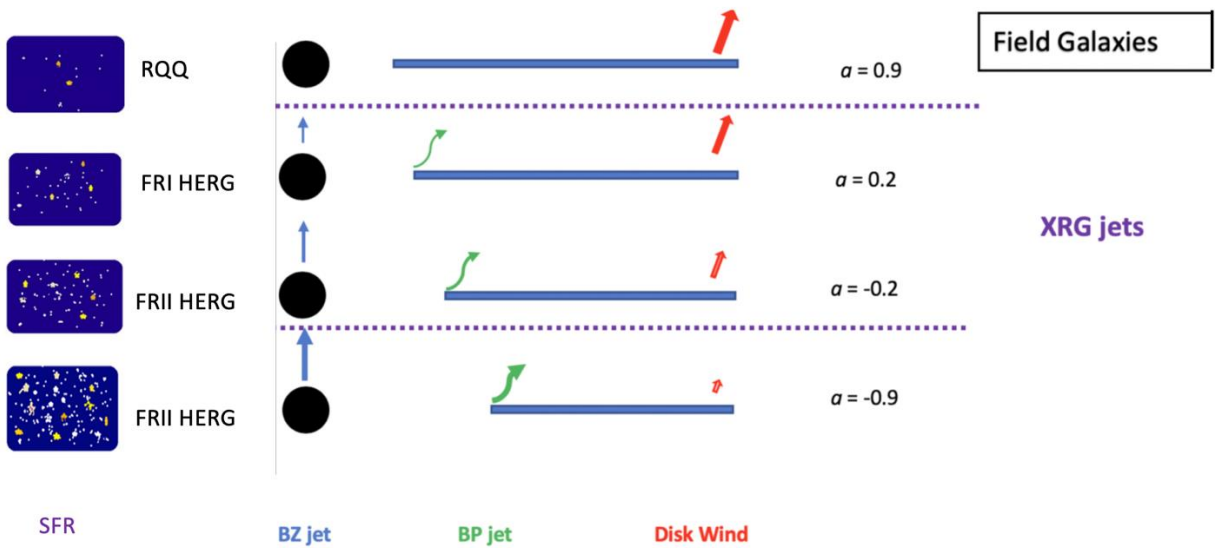

Figure 2: The origin and evolution of FRI quasars formed in mergers of isolated field galaxies with smaller black holes. Within the dashed purple lines, are the two XRG jets separated in time by about 5 million years.

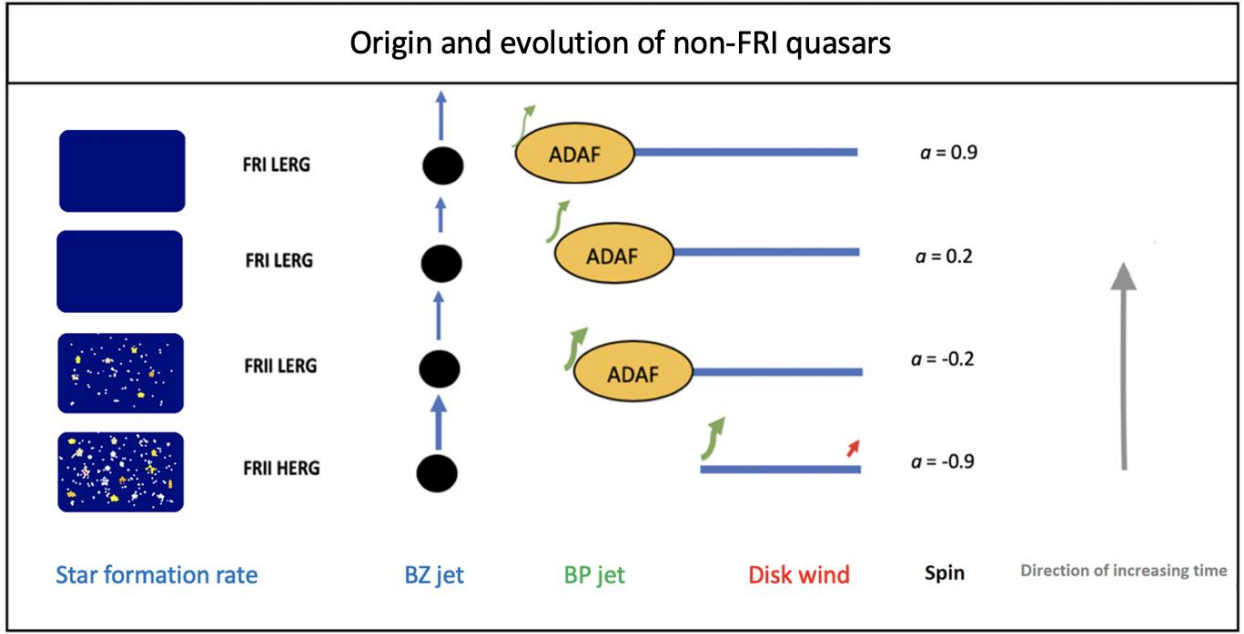

Figure 3: Evolution of jetted AGN in rich environments (from Garofalo & Mountrichas 2022). Such environments suffer an absence of FRI quasars. Despite the presence of an FRI jet, accretion is no longer in cold mode and thus the excitation level is low. Hence, no FRI HERG is formed. LERG = low excitation radio galaxy.

3. Final Thoughts

We have explored the parameter space behind the unusual characteristics of H1821+643 using three figures designed to capture the relevant features of the gap paradigm (Garofalo, Evans & Sambruna 2010). Figures 1-3 also give us a sense of why low excitation accretion is not a characteristic of accreting black holes that potentially experience jet suppression. In the subclass of isolated field galaxies that experienced a merger that led to cold gas around a black hole in counterrotation, the weak jet feedback on average fails to change the accretion state (Figure 2). In Figure 1, we see the same phenomenon in mergers or AGN fed by secular processes such as in spiral galaxies. As the environment becomes denser, instead, we see a characteristic change in the accretion state from a thin disk to an ADAF (Figure 3). Here we see an explanation/prediction as to why excitation class tends to be high at higher redshift in all environments, whereas at lower redshift we see a switch to lower excitation in denser cluster environments. These ideas allow us to include the recently discovered class of Γ-narrow line Seyfert 1 galaxies (Γ-NLS1; Foschini et al 2015) as FRI quasars triggered by secular processes in disk-like galaxies. As is for H1821+643, the prediction is that Γ-NLS1 harbor black holes whose spins are between 0.2 and 0.7. NLS1 galaxies would then fit in as the non-jetted version of such objects, which implies their black holes are spinning in the range $a < 0.1$ or $0.7 < a$. If the AGN reservoir of cold gas is sufficient, we also see why these corotating black holes tend to live longest in non-jetted phases. This is because Eddington-limited spin up from zero spin takes order $10^8$ years but once the 0.7 spin threshold is crossed (Garofalo & Singh 2016), the AGN persists in a radio quiet or jetless state. This has been estimated from observations to be up to $10^9$ years (Garofalo, Webster & Bishop 2020 and references therein). We end with the prediction based on the assumed correctness of the recent spin measurement of 0.62 from Sisk-Reynes et al (2022), that H1821+643 is within $10^7$ years of becoming a true radio quiet AGN.

Acknowledgments

CBS is supported by the National Natural Science Foundation of China under grant no. 12073021.